\documentclass[conference]{IEEEtran}
\IEEEoverridecommandlockouts

\usepackage[T1]{fontenc}
\usepackage{booktabs}
\usepackage{tabularx}
\usepackage{makecell}
\usepackage{float}
\usepackage{dblfloatfix}
\usepackage{placeins}
\usepackage{balance}
\usepackage{algorithm}
\usepackage{algpseudocode}

\usepackage{tikz,xcolor}

\DeclareRobustCommand\encircle[1]{%
  \tikz[baseline=(c.base)]{
    \node[
      shape=circle,
      draw,
      fill=white,
      inner sep=0.16ex,    
      minimum size=1.9ex,  
      line width=0.35pt
    ] (c) {\fontsize{7.5}{7.5}\selectfont #1};
  }%
}

\usepackage{cite}
\usepackage{amsmath,amssymb,amsfonts}
\usepackage{graphicx}
\usepackage{textcomp}
\usepackage{xcolor}

\usepackage[a4paper, total={184mm,239mm}]{geometry}

\def\BibTeX{{\rm B\kern-.05em{\sc i\kern-.025em b}\kern-.08em
    T\kern-.1667em\lower.7ex\hbox{E}\kern-.125emX}}

\begin{document}

\title{RAPID-Graph: Recursive All-Pairs Shortest Paths Using Processing-in-Memory for Dynamic Programming on Graphs}

\author{
    \IEEEauthorblockN{
        Yanru Chen\textsuperscript{*1}, 
        Zheyu Li\textsuperscript{*1}, 
        Keming Fan\textsuperscript{1}, 
        Runyang Tian\textsuperscript{1}, 
        John Hsu\textsuperscript{1}, \\
        Weihong Xu\textsuperscript{2}, 
        Minxuan Zhou\textsuperscript{3}, 
        Tajana Šimunić Rosing\textsuperscript{1}
    }

    \IEEEauthorblockA{
        \textsuperscript{1}University of California San Diego, La Jolla, CA, USA\\
    }

    \IEEEauthorblockA{
        \textsuperscript{2}Ecole Polytechnique Fédérale de Lausanne, Lausanne, Switzerland\\
    }

    \IEEEauthorblockA{
        \textsuperscript{3}Illinois Institute of Technology, Chicago, IL, USA\\
    }
    
    \IEEEauthorblockA{
        \{yac054, zhl178, k4fan, r3tian, joh048, tajana\}@ucsd.edu; weihong.xu@epfl.ch; mzhou26@iit.edu
    }

}

\IEEEaftertitletext{\vspace{-1cm}}

\maketitle

\begin{abstract}

All-pairs shortest paths (APSP) remains a major bottleneck for large-scale graph analytics, as data movement with cubic complexity overwhelms the bandwidth of conventional memory hierarchies. In this work, we propose RAPID-Graph to address this challenge through a co-designed processing-in-memory (PIM) system that integrates algorithm, architecture, and device-level optimizations. At the algorithm level, we introduce a recursion-aware partitioner that enables an exact APSP computation by decomposing graphs into vertex tiles to reduce data dependency, such that both Floyd-Warshall and Min-Plus kernels execute fully in-place within digital PIM arrays. At the architecture and device levels, we design a 2.5D PIM stack integrating two phase-change memory compute dies, a logic die, and high-bandwidth scratchpad memory within a unified advanced package. An external non-volatile storage stack stores large APSP results persistently. The design achieves both tile-level and unit-level parallel processing to sustain high throughput. On the 2.45M-node OGBN-Products dataset, RAPID-Graph is $5.8\times$ faster and $1\,186\times$ more energy efficient than state-of-the-art GPU clusters, while exceeding prior PIM accelerators by $8.3\times$ in speed and $104\times$ in efficiency. It further delivers up to $42.8\times$ speedup and $392\times$ energy savings over an NVIDIA H100 GPU.

\end{abstract}


\section{Introduction}

Graphs serve as the backbone of workflows spanning urban planning and transportation~\cite{putriani2024comparison,aktacs2024speeding}, social and commercial analytics~\cite{ma2024using,simas2021distance}, and autonomous LiDAR navigation~\cite{wan2021survey}. Central to these applications, all-pairs shortest paths (APSP) provides a fundamental computational core, transforming graph connectivity into actionable insights. However, exact APSP computation remains computationally prohibitive due to inherent algorithmic complexity. The classic Floyd-Warshall (FW) algorithm requires \(O(n^3)\) time and \(O(n^2)\) space~\cite{floyd1962algorithm}, while the repeated Dijkstra algorithm exhibits super-quadratic complexity with poor memory locality~\cite{dijkstra2022note}.

Recent GPU-based APSP accelerators achieve high throughput by launching thousands of parallel threads, but require massive hardware resources, intensive interconnect communication, and high energy consumption~\cite{shi2018graph,xie2025gpu}. Partitioned-APSP computes APSP for a 2M-vertex graph in approximately 30 minutes but requires 128 GPUs with extensive DRAM reliance~\cite{djidjev2015all}; Co-ParallelFW achieves 8.1\,PFLOP/s but requires complex coordination among 4,608 GPUs~\cite{sao2021scalable}; Fast-APSP scales up to 11.5M vertices utilizing 2\,048 GPUs with compressed sparse row (CSR)-tile layouts and complex min-plus (MP) reductions, yet remains bottlenecked by inter-GPU communication overhead~\cite{yang2023fast}.

Processing-in-memory (PIM) reduces data movement by executing logic inside memory arrays. Prior PIM architectures validated through graph traversals: GraphR accelerates BFS through in-situ SpMV on ReRAM crossbars~\cite{8327035}; GraphPIM enables atomic rank-level SSSP updates using DRAM bit-line logic~\cite{7920847}; and Tesseract manages vertex-centric execution via remote procedure calls across HMC stacks~\cite{ahn2015scalable}. Similarly, Temporal State Machines~\cite{madhavan2021temporal} accelerate Dijkstra SSSP by executing tropical algebra operations on time-coded wavefronts, achieving an edge traversal rate of 10 giga-edge traversals per second with a memristive temporal processor. Later systems such as GraphP~\cite{zhang2018graphp}, GraphH~\cite{sun2017graphh}, and GraphQ~\cite{zhuo2019graphq} advanced the state-of-the-art (SOTA) with improved inter-cube communication and scheduling, yet they still focus on sparse traversals and depend on atomic operations over lightweight data. Exact APSP instead requires dense $O(n^2)$ matrices and heavy MP merges that quickly overwhelm on-chip SRAM and local memory bandwidth. No existing PIM design addresses this quadratic storage requirement or provides efficient in-memory MP reductions needed for APSP dynamic programming (DP).

To support dense APSP workloads, PIM architectures must combine computational innovation with suitable memory technology, as the cubic time and quadratic space complexity of APSP demands high-capacity, high-speed, and energy-efficient in-place computation. Filamentary RRAM enables sub-$0.1$\,pJ writes, but 4-bit HfO$_2$ stacks~\cite{he2023ti} need $\sim50$\,\textmu s pulses and retain data for only $\sim$$10^{5}$\,s at $85$$^\circ$C, limiting throughput and longevity. Spin-orbit-torque MRAM~\cite{nguyen2025low} writes in $0.35$\,ns at $156$\,fJ/bit, but its $6-8$\,$F^2$ MTJ-transistor stack has lowers density. By contrast, SiTe$_x$ filamentary phase-change memory (PCM)~\cite{park2024phase} switches in $150$ ns and $20$ ns at $\sim10$\,pJ, demonstrating a better energy-speed balance. Sb\textsubscript{2}Te\textsubscript{3}/Ge\textsubscript{4}Sb\textsubscript{6}Te\textsubscript{7} 40 nm PCM~\cite{wu2024novel} offers $\sim20$ ns switching, $1.5$ pJ reset, $10^8$ endurance, and $10^5$ h retention at $83$°C, with a $150\times$ $R_\mathrm{on}/R_\mathrm{off}$ ratio. Therefore, this work adopts the 40 nm Sb\textsubscript{2}Te\textsubscript{3}/Ge\textsubscript{4}Sb\textsubscript{6}Te\textsubscript{7} PCM technology, leveraging its flash-like endurance, SRAM-class speed, high density, and pJ-level energy to support extensive parallelism and memory bandwidth required by APSP.

To break through the fundamental memory and power walls that create critical performance bottlenecks in exact APSP computation, we introduce RAPID-Graph, a novel software-hardware co-designed PIM system. Our approach overcomes the communication limits of multi-GPU clusters and the dense-workload constraints of prior PIM architectures. Our core contributions are as follows:

\begin{itemize}
\item Recursive in-memory APSP dataflow. We design an algorithm that recursively decomposes the graph into subproblems sized for a single PIM tile, enabling massively parallel FW execution entirely in PCM arrays and eliminating off-chip data movement.

\item Heterogeneous 2.5D PIM architecture. We build a tailored architecture for this dataflow, integrating PCM compute dies with a logic die via UCIe, supported by an on-package HBM3 for hot submatrices and external FeNAND for bulk storage, forming a balanced memory hierarchy.

\item SOTA performance and efficiency. RAPID-Graph delivers up to $1\,061$× speedup and $7\,208$× energy efficiency over CPU. On the OGBN-Products dataset (2.5M nodes), it outperforms GPU clusters by $5.8$× in runtime and $1\,186$× in energy efficiency.
\end{itemize}

\section{Background}
\label{sec:background}

\subsection{Graphs}

\begin{figure}[t]
    \centering
    \includegraphics[width=0.95\linewidth]{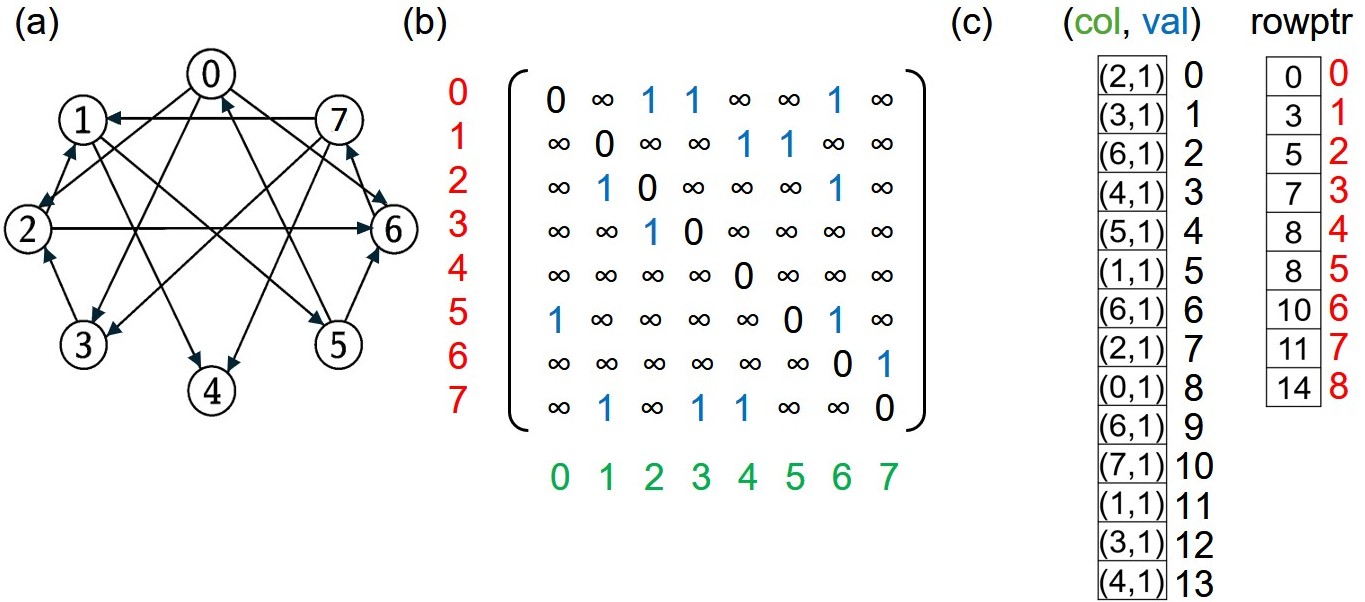}
    \caption{Graph representations (a) Graph topology (b) Adjacent matrix (c) CSR}
    \label{fig:csr_dense_contrast}
\end{figure}

A graph $G=(V,E,w)$ can be stored as an adjacency matrix or CSR. Computation is most convenient on its dense adjacency matrix $A$, defined by $A_{ij}=w(i,j)$ for $(i,j)\in E$ and $+\infty$ otherwise, enabling regular dataflow for FW and MP kernels. For storage, we adopt CSR, which records only the nonzero weights together with row boundaries, reducing space from $n^{2}$ to $|E|$. Fig.~\ref{fig:csr_dense_contrast} illustrates an 8-vertex toy graph in three views: (a) the original topology, (b) the dense matrix with finite weights marked, and (c) the compressed CSR ${\texttt{rowptr}, \texttt{col}, \texttt{val}}$, demonstrating space savings when $|E|\ll n^2$.

\subsection{APSP Algorithms}
\subsubsection{Classic FW DP}

The FW algorithm~\cite{floyd1962algorithm} solves the APSP problem on a weighted graph \( G=(V,E,w) \) via an in-place dynamic program over an \( n \times n \) distance matrix \( D \). $D[i][j]$ is set to the weight of edge $(i,j)$, or $\infty$ if no edge exists. The algorithm updates all entries by checking whether paths through vertex $k$ yield shorter distances by:
\[
D[i][j] = \min\left(D[i][j],\; D[i][k] + D[k][j]\right), k \in [1, n].
\]
After $n$ iterations, $D$ contains the exact shortest path lengths between all vertex pairs. The algorithm runs in \(O(n^3)\) time and \(O(n^2)\) space, following a computation pattern equal to a dense outer product of row \( i \) and column \( k \) against column \( j \).


\subsubsection{Partitioned APSP}

\begin{figure}[t]
    \centering
    \includegraphics[width=0.8\linewidth]{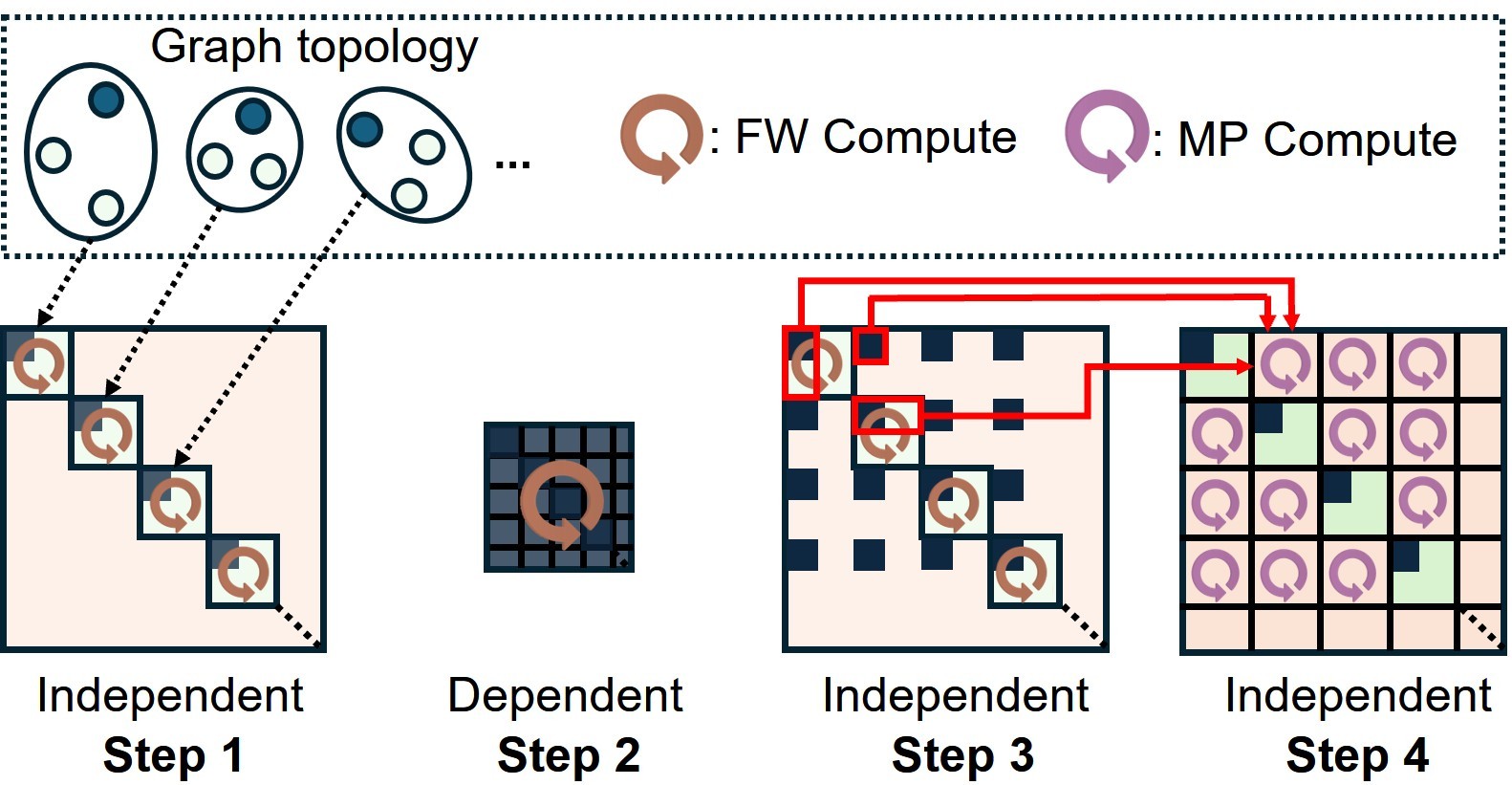}
    \caption{Illustration of partition APSP}
    \label{fig:partitioned-apsp}
\end{figure}

Our recursive partition APSP extends the four-stage scheme by~\cite{djidjev2015all}, which reshapes FW into tile-local semiring operations, aligning with PIM’s in-place parallelism. The algorithm is illustrated in Fig.\ref{fig:partitioned-apsp} and summarized in Algorithm~\ref{alg:partitioned-apsp}.
Graph preprocessing runs on host CPU, where a weighted graph \(G\) is partitioned into components \(C_1,\dots,C_k\) and their boundary set \(B\) via a $k$-way \textsc{Metis}~\cite{karypis1998multilevelk}. Within each component, a boundary vertex has an edge connecting to another component, while an internal vertex only has edges to vertices within its own component. For computational efficiency, boundary vertices are reordered before internal vertices.

\textbf{Step 1: Local APSP.} Each component independently runs FW to fill its intra-component distance matrix \(d_\text{intra}\); all passes execute in parallel and scale linearly.

\textbf{Step 2: Boundary-graph APSP.} All boundary vertices form a reduced graph \(G_B\), with edges comprising: (i) cross-component edges from \(G\), and (ii) virtual edges within components weighted by \(d_\text{intra}\). A single FW run computes the boundary distance matrix \(d_B\). As this step involves dense \(O(|B|^{3})\) work, it becomes the primary bottleneck when \(|B|\) nears hardware limits. Section~\ref{sec:software} addresses this bottleneck through recursive partitioning.

\textbf{Step 3: Boundary injection.} Each component copies the relevant rows and columns of \(d_B\) into its local matrix and re-runs FW once, propagating inter-component shortcuts.

\textbf{Step 4: Cross-component merge.} An MP merge combines (i) source to boundary, (ii) boundary to boundary, and (iii) boundary to destination paths, producing the final cross-component distances \(d_\text{cross}\) and thus completing global APSP.

\begin{algorithm}[t]
\scriptsize
\caption{Partition APSP Pseudocode}
\label{alg:partitioned-apsp}
\begin{algorithmic}[1]
\State Partition \(G\) into \(k\) components \(C_1, \dots, C_k\) via \textsc{Metis}~\cite{karypis1998multilevelk}
\State \textbf{for} \(i = 1\) \textbf{to} \(k\) \textbf{do} \hfill \Comment{Step 1}
\State \hspace{1em} \textsc{FW}\((C_i)\)
\State \(G_B \leftarrow \textsc{extractBoundaryGraph}(G)\) \hfill \Comment{Step 2}
\State \textsc{FW}\((G_B)\)
\State \textbf{for} \(i = 1\) \textbf{to} \(k\) \textbf{do} \hfill \Comment{Step 3 with injected \(d_B\)}
\State \hspace{1em} \textsc{FW}\((C_i)\)
\State \textbf{for} \(i = 1\) \textbf{to} \(k\) \textbf{do} \hfill \Comment{Step 4}
\State \hspace{1em} \textbf{for} \(j = 1\) \textbf{to} \(k\) \textbf{do}
\State \hspace{2em} \textsc{MinPlusMerge}\((C_i, C_j)\)
\State \textbf{return} global distance matrix
\end{algorithmic}
\end{algorithm}

\subsection{Phase Change Memory}

PCM~\cite{wong2010phase} encodes data by switching a chalcogenide cell between high-resistance amorphous state (reset, 0) and low-resistance crystalline state (set, 1) via Joule heating. Cells form 1T1R cross-point arrays whose wordlines and bitlines are managed by selection transistors. PCM offers fast random reads, low leakage, and non-volatility, making it ideal for in-place parallel updates and persistent storage in our PIM system.






Our PCM array supports bit-serial logic primitives inspired by FELIX~\cite{gupta2018felix}. Single-cycle NOR, NOT, NAND, minority, OR and 2-cycle XOR operations execute natively in crossbar memory via voltage-controlled switching. These primitives enable efficient composition of more complex functions: 1-bit addition and min-comparison are implemented via FELIX~\cite{gupta2018felix} primitives. Addition computes the sum bit $S = A \oplus B \oplus C_{in}$, and the carry-out $C_{out} = Maj(A, B, C_{in})$. Min-comparison uses bit-serial subtraction $S = A \oplus (\neg B) \oplus 1$, where most significant bit of $S$ is the sign bit gating selective updates.

\section{RAPID-Graph Software-Hardware Co-Design}
\label{sec:systemarch}


\subsection{Recursive Partitioned APSP}
\label{sec:software}
\begin{figure}[t]
    \centering
    \includegraphics[width=1\linewidth]{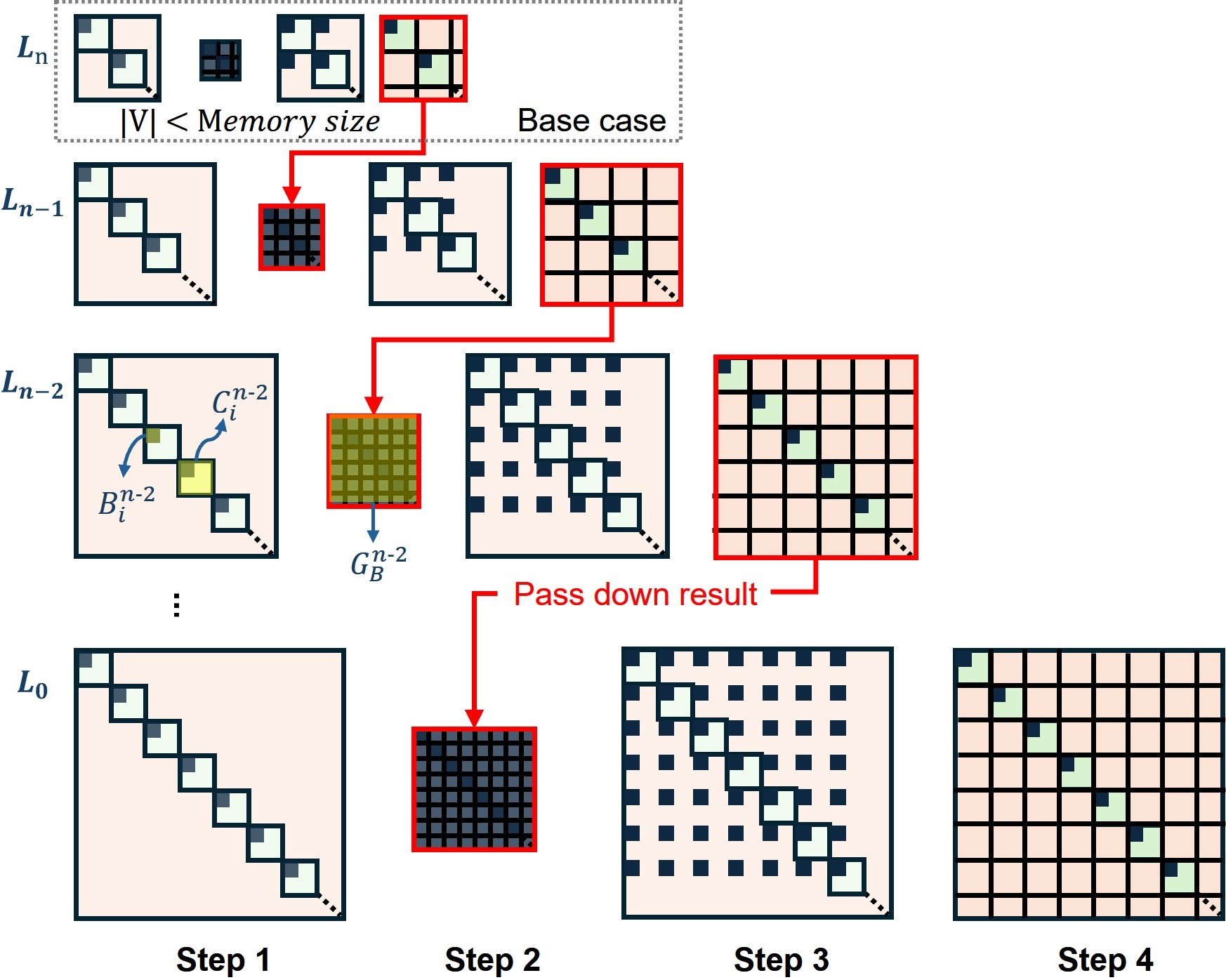}
    \caption{Illustration of recursive partitioned APSP}
    \label{fig:recursion-apsp-flow}
\end{figure}

To efficiently scale APSP computation on large graphs, we design a recursive partitioning strategy that enables fully independent subgraphs sized to fit within PIM tile limits. Accordingly, we partition each component at $|V| \leq 1024$, matching practical array dimensions per PCM tile and the maximum parallelism achievable with dense, high-yield fabrication. The input graph $G=(V,E,w)$, with vertex set $V$, edge set $E$, and non-negative weights $w$, is first partitioned by METIS~\cite{karypis1998multilevelk} into base-level components $C_1^{(0)}, \dots, C_k^{(0)}$. Each component $C_i^{(0)}$ contains internal vertices and boundary vertices, where boundary vertices connect to other components. We extract boundary vertices from ${C_i^{(0)}}$ to construct the level-0 boundary graph $G_B^{(0)}$. If boundary graph $G_B^{(\ell)}$ at level $\ell$ exceeds the 1024-vertex tile limit, we recursively partition it, creating a coarser graph $G_B^{(\ell+1)}$. This continues until $|V(G_B^{(n)})|\leq1024$, ensuring all graphs fit entirely within tiles. Table~\ref{tab:variables} summarizes key variables. At each recursion level, APSP is computed locally within components and boundary graphs, propagating distances back into components and performing cross-component updates via MP products. Fig.~\ref{fig:recursion-apsp-flow} illustrates this process. After recursive partitioning each memory array holds one dense distance block with \(N \le 1024\) vertices. To maximize parallelism during the FW updates, we adopt a specialized data remapping strategy. This strategy logically separates the current pivot row and column (the \texttt{Panel\_Row} and \texttt{Panel\_Col}) from the rest of the distance matrix (the \texttt{Main\_Block}). This layout allows the FW die to perform massively parallel updates on the \texttt{Main\_Block}. The detailed mapping and scheduling schemes for this process are described in Section~\ref{sec: mapping}.

\begin{table}[t]
\caption{Key Variables in Recursive Partitioned APSP}
\label{tab:variables}
\scriptsize
\renewcommand{\arraystretch}{0.9}
\setlength{\tabcolsep}{3pt}
\centering
\begin{tabular}{|>{\raggedright\arraybackslash}p{2.2cm}|>{\raggedright\arraybackslash}p{6.1cm}|}
\hline
\textbf{Variable} & \textbf{Description} \\
\hline
$G = (V,E,w)$ & Graph with vertices $V$, edges $E$, edge weights $w$ \\
$C_i^{(\ell)}$ & Component at level $\ell$ ($\ell$=1,...,n and i=1,...,k) \\
$B_i^{(\ell)}$ & Boundary vertex set of component $C_i^{(\ell)}$ \\
$G_B^{(\ell)}$ & Level-$\ell$ boundary graphs \\
$\mathrm{DB}^{(\ell)}$ & Boundary-to-boundary distance matrix at level $\ell$ \\
$D_C$ & Intra-component APSP distance of component $C$ \\
$D_{C_1}[m,n]$ & Cross-component distance from $m\in C_1$ to $n\in C_2$\\
\hline
\end{tabular}
\end{table}

Algorithm~\ref{alg:recursive-apsp} summarizes the recursive APSP procedure across hierarchy levels. It follows the same four steps as Algorithm~\ref{alg:partitioned-apsp}, but operates bottom-up: starting from base-level partitions, each level computes local APSP and propagates boundary summaries upward. This structure maximizes tile-level parallelism and fits within constrained PIM resources without requiring global synchronization.

\begin{algorithm}[t]
\scriptsize
\caption{Recursive Partition APSP Pseudocode}
\label{alg:recursive-apsp}
\begin{algorithmic}[1]
\State \textbf{for} \(\ell = n\) down to \(0\) \textbf{do}
\State \hspace{1em} \textbf{parallel for} \(C\) in levels[\(\ell\)] \textbf{do} \hfill \Comment{Step 1}
\State \hspace{2em} \(D_C \gets \text{FloydWarshall}(C)\)
\State \hspace{2em} \(B_C \gets \text{find\_boundary}(C)\)
\State \hspace{2em} \textbf{if} DB\_prev == \(\varnothing\): \(DB_C \gets \text{restrict}(D_C, B_C)\)
\State \hspace{1em} \textbf{if} DB\_prev == \(\varnothing\): \hfill \Comment{Step 2}
\State \hspace{2em} \(G_B \gets \text{build\_boundary\_graph}(\{DB_C \text{ for all } C\})\)
\State \hspace{2em} DB\_prev \(\gets \text{FloydWarshall}(G_B)\)
\State \hspace{1em} \textbf{parallel for} \(C\) in levels[\(\ell\)] \textbf{do} \hfill \Comment{Step 3}
\State \hspace{2em} \(D_C \gets \text{inject}(DB\_prev, B_C)\)
\State \hspace{1em} \textbf{parallel for} \((C_1, C_2)\) in levels[\(\ell\)] \textbf{do} \hfill \Comment{Step 4}
\State \hspace{2em} \textbf{for} \(m\) in \(B_{C_1}\), \(n\) in \(B_{C_2}\) \textbf{do}
\State \hspace{2em} \(D_{C_1}[m,n] \gets \min\limits_{\substack{i \in B_{C_1} \\ j \in B_{C_2}}} \left( D_{C_1}[m,i] + \text{DB\_prev}[i,j] + D_{C_2}[j,n] \right)\)
\State \hspace{1em} DB\_prev \(\gets \text{merge}(\{\text{restrict}(D_C, B_C)\})\) 
\State \textbf{return} DB\_prev
\end{algorithmic}
\end{algorithm}

\subsection{Heterogeneous Architecture Overview}
\label{sec:heterogeneousoverview}

\begin{figure}[t]
    \centering
    \includegraphics[width=1\linewidth]{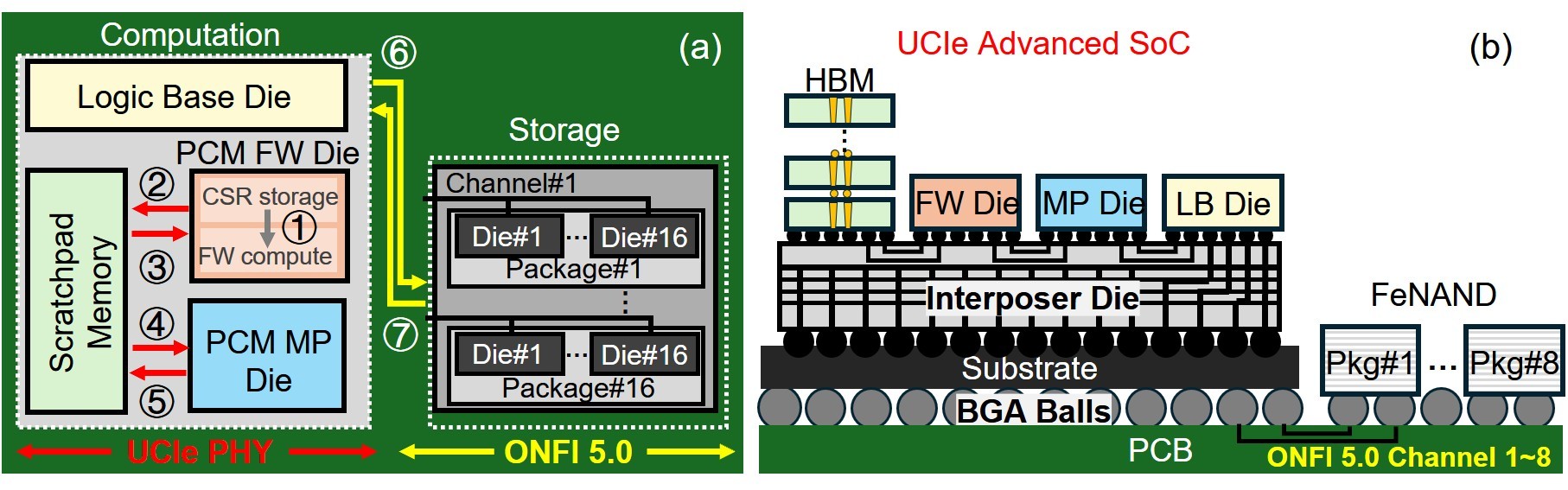}
    \caption{RAPID-Graph heterogeneous PIM system 
        (a) System-level dataflow of UCIe-based multi-die SoC and off-package FeNAND storage, annotated with computation-storage dataflow
        (b) 2.5D advanced packaging cross-section}
    \label{fig:multi-die-architecture}
\end{figure}

To scale APSP, we build a heterogeneous multi-die architecture that integrates high-bandwidth in-memory compute with large nonvolatile storage, as shown in Fig.\ref{fig:multi-die-architecture}(a). We co-package the PCM-FW die, PCM-MP die, logic base die, and HBM3 using a 2.5D silicon interposer with a UCIe v1.0 raw interface~\cite{lin202536}, delivering 64 full-duplex lanes at 32 Gb/s each, as illustrated in Fig.~\ref{fig:multi-die-architecture}(b). Off-package FeNAND is an external storage array mounted on the PCB and connected via ONFI 5.1 ×16 channels. It delivers several key advantages over conventional charge-trap NAND, including higher capacity, lower program voltage, faster access, and a shorter string length~\cite{yun2024optimization}. This multi-die design maps the recursive APSP algorithm’s FW and MP kernels onto dedicated PIM compute dies. Fine-grained partitioning, coupled through the high-bandwidth UCIe link, creates a tailored dataflow that eases the memory and communication bottlenecks of large-scale APSP.


Inside the package, the system components are organized by function. The logic base die serves as the central controller, managing system-wide dataflow and leveraging its dual stream engines for high-throughput CSR-to-dense format conversion. In-memory computation is executed on two specialized 2\,GB PCM compute dies: the FW die is tailored for intra-component updates, while the MP die accelerates cross-partition merges. This processing core is supported by a tiered memory hierarchy, where a 16\,GB HBM3 stack provides high-bandwidth scratchpad buffering, and a 16\,TB FeNAND~\cite{10873526} array offers dense persistent storage for boundary data and final results in compressed CSR format.



This purpose-built hardware enables a multi-stage dataflow, orchestrated by the logic die to execute the recursive APSP algorithm across the memory hierarchy. Fig.~\ref{fig:multi-die-architecture}(a) summarizes the recursive APSP dataflow across the memory hierarchy:

\encircle{1} CSR-partitioned components are streamed from the cold-storage region of the PCM-FW die into its compute region and expanded into dense matrices.
\encircle{2} The PCM-FW die performs local FW updates, and results are streamed back to HBM3 in row-wise segments.
\encircle{3} HBM3 (i) extracts boundary nodes and edges to construct a boundary graph for the next FW iteration, and (ii) prefetches next intra-component FW blocks for pipelined execution.
\encircle{4} The PCM-MP die fetches submatrices from HBM3 to perform cross-partition MP merges.
\encircle{5} HBM3 synchronizes boundary data across partitions to prepare for the next boundary-aware FW round.
\encircle{6} (i) Updated boundary matrices are stored in dense format in FeNAND, whereas final cross-partition APSP results from PCM-MP and (ii) intra-component APSP from PCM-FW are compressed to CSR format before writing to FeNAND.
\encircle{7} The PCM-MP die fetches interleaved boundary matrices from FeNAND to complete final cross-partition merges.


\subsection{Hardware Design and Implementation}
\label{sec:pcm-die-arch}

\begin{figure}[t]
    \centering
    \includegraphics[width=1\linewidth]{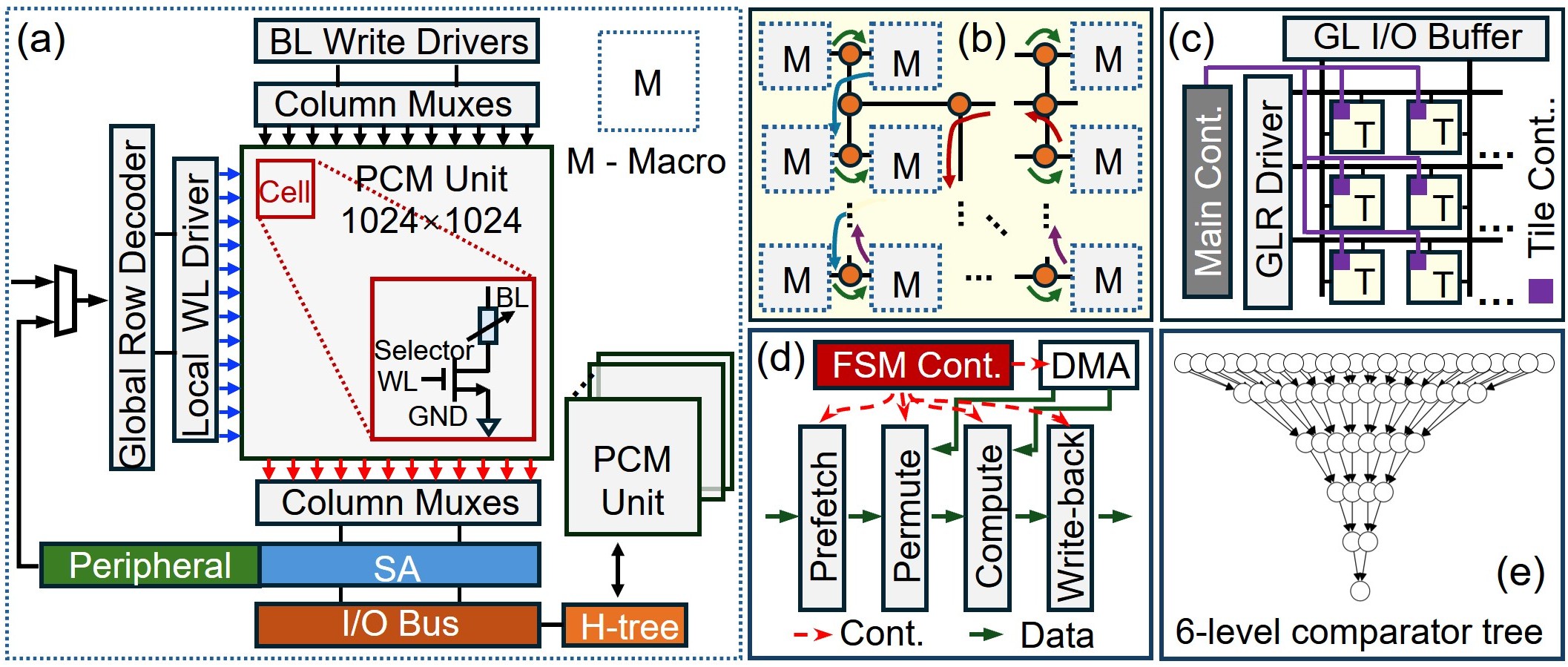}
    \caption{PCM die detail architecture (a) Macro - M (b) Tile - T (c) Die (d) Permutation unit for PCM-FW (e) 6-level min comparator tree for PCM-MP}
    \label{fig:pim_arch}
\end{figure}

Both PCM-FW and PCM-MP dies use the same 1T1R SLC PCM cell technology~\cite{wu2024novel}, organized into $1024 \times 1024$ units. The parameters for the  SLC cells are listed in Table~\ref{tab:pcm-parameters}. Each unit, together with its peripheral circuits, forms a macro, as shown in Fig.\ref{fig:pim_arch}(a). Each tile contains 130 parallel units connected via an H-tree interconnect~\cite{7551379} for efficient data exchange (Fig.~\ref{fig:pim_arch}(b)). The design ensures full crossbar activity without idle cycles. Same-colored arrows represent concurrent unit-to-unit transfers.
Tiles operate concurrently, each with a local controller linked to a global main controller for coordinated execution (Fig.~\ref{fig:pim_arch}(c)). In-memory operations are triggered by row-segment broadcasts, with each tile controller managing parallel processing across its units. In addition to shared peripheral logic, each PCM die integrates specialized circuits for its role. The PCM-FW tile includes a permutation unit for rearranging data blocks, while the PCM-MP tile integrates a 32-bit 6-level min-comparator tree for efficient MP reductions.

\begin{table}[t]
\caption{Key Parameters of Sb\textsubscript{2}Te\textsubscript{3}/Ge\textsubscript{4}Sb\textsubscript{6}Te\textsubscript{7} SLC PCM}
\label{tab:pcm-parameters}
\scriptsize
\renewcommand{\arraystretch}{0.9}
\setlength{\tabcolsep}{3pt}
\centering
\begin{tabular}{|>{\raggedright\arraybackslash}p{3.2cm}|>{\raggedright\arraybackslash}p{3.5cm}|}
\hline
\textbf{Parameter} & \textbf{Value} \\
\hline
Reset pulse & 40\,$\mu$A @ 0.70\,V \\
Programming energy & $\approx$ 0.56\,pJ \\
Resistance (LRS) & $\approx$ 30\,k$\Omega$, 150$\times$ on/off ratio \\
Set/reset time & 20\,ns / 20\,ns \\
Clock cycle & 2\,ns (500\,MHz) \\
\hline
\end{tabular}
\end{table}

\textbf{PCM-FW Permutation Unit.} 
The PCM-FW die includes a dedicated permutation macro that locally rearranges data blocks, avoiding off-die data movement. As shown in Fig.~\ref{fig:pim_arch}(d), the permutation macro comprises:

\begin{enumerate}
  \item a 32-row burst row-buffer controller,
  \item a reorder buffer for panel masking and block pruning,
  \item a lightweight on-tile DMA engine (1-cycle read, 10-cycle write) with address remapper, and
  \item a four-stage FSM pipeline \small(Prefetch $\!\rightarrow$ Permute $\!\rightarrow$ Compute $\!\rightarrow$ Write-back\small),
\end{enumerate}
so that data movement overlaps computation.  
The permutation unit packs \texttt{Panel\_Row} and mirrored \texttt{Panel\_Col} into 32-row windows for coalesced bursts without H-tree stalls, while a \texttt{prefetch} buffer hides DMA latency and skip futile writes, sustaining near-peak occupancy and reducing wear.

\textbf{PCM-MP Min-Compare \& Update.} 
Each unit integrates a pipelined comparator tree (Fig.~\ref{fig:pim_arch}(e)) that reduces 1024 32-bit inputs to one 32-bit minimum in following periods: A 1024 × 32-bit row is streamed into the buffer in 1 cycle. Thirty-two parallel five-level carry look-ahead (CLA) trees extract sign bits and 5-bit indices for block minima in 6 cycles. A second five-level tree reduces these to a global minimum in another 6 cycles, totaling 13 cycles per row. The final sign-bit mask gates PCM writes, updating only entries with smaller values. Two staging buffers hold operands across adds while $DB[i,j]$ streams, enabling one 1024-wide vector per cycle. The reduction tree outputs an update mask that enables compare-and-swap selective writes, avoiding read–modify–write and lowering energy and wear.


\subsection{Mapping and Scheduling Schemes} 
\label{sec: mapping}

\begin{figure}[t]
    \centering
    \includegraphics[width=1\linewidth]{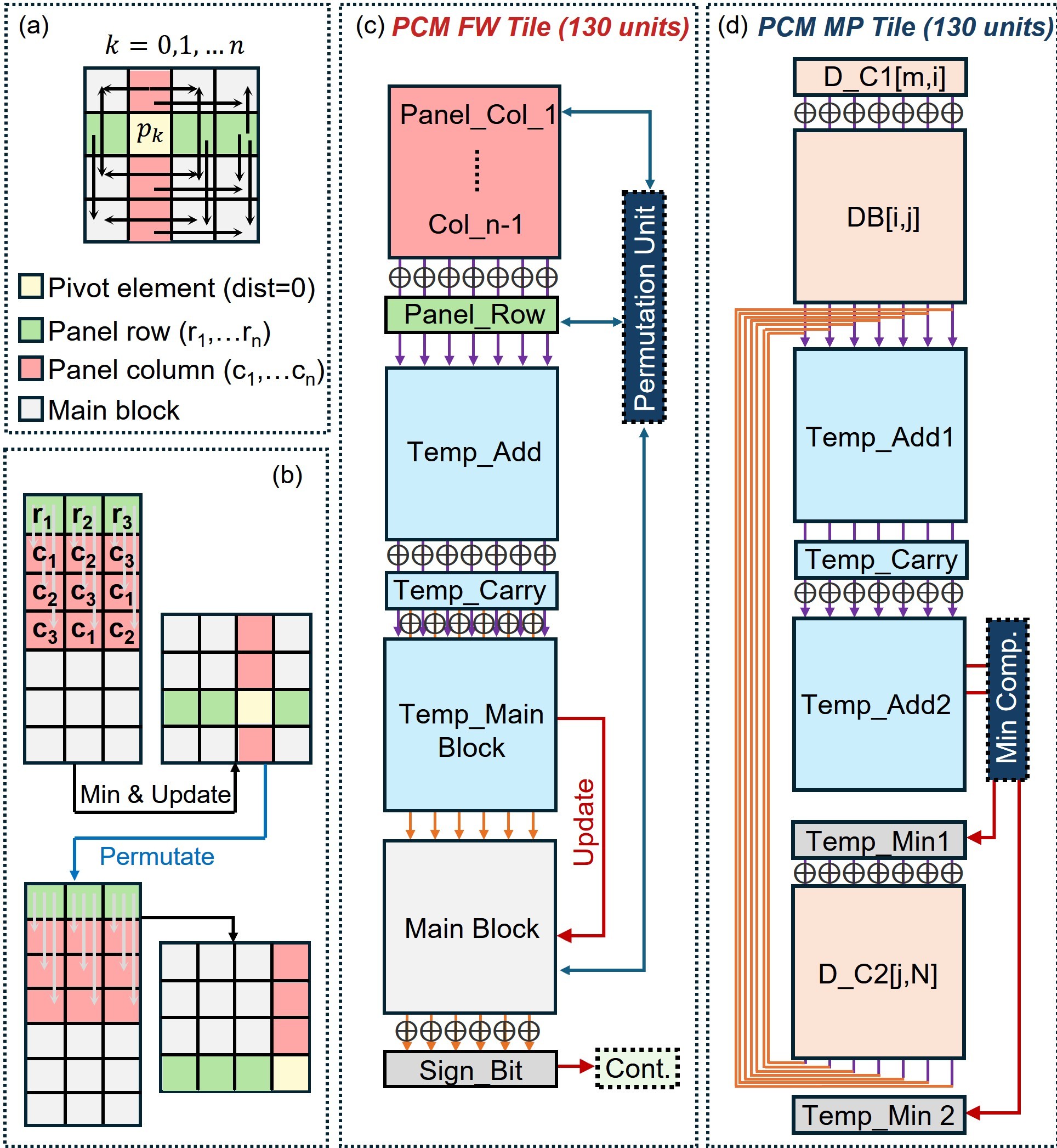}
    \caption{SW/HW mapping for recursive partitioned APSP (a) FW illustration (b) FW remapping, every step includes add, min update and permutation (c) PCM-FW tile performing FW (d) PCM-MP tile performing two-stage MP}
    \label{fig:Mapping}
\end{figure}

To execute FW and MP efficiently in PCM tiles, we co-optimize software layout and hardware design through tailored intra-tile mapping and scheduling. Fig.~\ref{fig:Mapping}(a) illustrates the FW DP update flow. Since diagonal pivot elements $p_k$ always have zero distance, their propagation along pivot row ($r_1$,...$r_n$) and column ($c_1$,...$c_n$) is omitted. Instead, row and column elements propagate into the main block to perform add and min operations, updating main block values with MP products. Each pivot element triggers one such update; the process repeats until all diagonal pivots $p_k$ ($k$=1,..,n) are processed. Fig.~\ref{fig:Mapping}(b) shows our remapping strategy that maximizes array parallelism. We extract the pivot row, copy its column beneath it, and form an $(n{-}1) \times (n{-}1)$ block; this lets all updates finish with one add and one min. The next pivot element is then permutated, and the process repeats.

Fig.~\ref{fig:Mapping}(c) shows the PCM-FW tile architecture, comprising 130 units. The pivot column elements are streamed into \textcolor[RGB]{210,100,100}{\texttt{Panel\_Col}[1:\,$n-1$]}, the pivot row into \textcolor[RGB]{60,140,60}{\texttt{Panel\_Row}}, and all other entries into \textcolor[RGB]{90,90,90}{\texttt{Main\_Block}}. Additions are implemented by FELIX~\cite{gupta2018felix}-style bit-serial units using \textcolor[RGB]{90,170,200}{\texttt{Temp\_Add}} and \textcolor[RGB]{90,170,200}{\texttt{Temp\_Carry}}. A FELIX~\cite{gupta2018felix} bit-serial subtraction compares the intermediate results in \textcolor[RGB]{90,170,200}{\texttt{Temp\_Main\_Block}} against \textcolor[RGB]{90,90,90}{\texttt{Main\_Block}}; the resulting sign bit gates selective writes back to \textcolor[RGB]{90,90,90}{\texttt{Main\_Block}}, completing one full main-block update. A dedicated permutation unit then reorders \textcolor[RGB]{60,140,60}{\texttt{Panel\_Row}}, \textcolor[RGB]{210,100,100}{\texttt{Panel\_Col}} and \textcolor[RGB]{90,90,90}{\texttt{Main\_Block}} for the next pivot iteration, as detailed in Section~\ref{sec:pcm-die-arch}. Fig.~\ref{fig:Mapping}(d) shows the PCM-MP tile architecture, comprising 130 units that execute a two-stage MP merge. In the first stage, a logical row $(1 \times 1024)$ from $C_1$ is loaded into \textcolor[RGB]{200,120,80}{$D_{C_1}[m,i]$}, while \textcolor[RGB]{200,120,80}{$DB[i,j]$} and the corresponding row \textcolor[RGB]{200,120,80}{$D_{C_2}[j,n]$}, representing paths from boundary $j$ to internal node $n$ in $C_2$, are presented in parallel. The merge proceeds via two successive MP operations: \textcolor[RGB]{200,120,80}{$D_{C_1}[m,i]$} + \textcolor[RGB]{200,120,80}{$DB[i,j]$} followed by $(\cdot) + $\textcolor[RGB]{200,120,80}{$D_{C_2}[j,n]$}, each performing a 1024-way Min reduction. Computation is performed using FELIX-style bit-serial adders (\textcolor[RGB]{90,170,200}{\texttt{Temp\_Add1}, \texttt{Temp\_Carry}, \texttt{Temp\_Add2}}), and a pipelined comparator tree reduces all candidates to a single minimum.



\section{Experimental Results and Analysis}
\label{sec:evaluation}

\subsection{Experimental Setup}

We benchmark RAPID-Graph on the real-world OGBN-products graph~\cite{chiang2019cluster} for fair SOTA comparison, and on synthetic Newman–Watts–Strogatz (NWS)~\cite{watts1998collective} and Erdős–Rényi (ER)~\cite{erdds1959random} graphs generated via NiemaGraphGen~\cite{moshiri2022niemagraphgen} for controlled scalability and topology analysis. NWS preserves dense intra-community but sparse inter-community links, while ER has uniformly random edges. These datasets reveal how graph topology impacts RAPID-Graph’s scalability.


We compare against CPU, GPU, SOTA PIM method Temporal PIM SSSP~\cite{madhavan2021temporal}, and SOTA GPU distributed methods including Partitioned APSP~\cite{djidjev2015all} and Co-Parallel APSP~\cite{sao2021scalable}. Since no SOTA PIM methods directly implement APSP, we estimate the performance of the Temporal PIM SSSP~\cite{madhavan2021temporal} to establish a comparable APSP PIM baseline, hereafter referred as PIM-APSP. Graph partitioning is performed using METIS 5.1.0~\cite{karypis1998multilevelk}. Consistent with prior work~\cite{djidjev2015all,sao2021scalable}, the METIS partitioning overhead is not included in our results as it is a preprocessing step. The baseline configurations are:

\begin{enumerate}
\item \textbf{CPU}: Intel i7-11700K (64 GB)
\item \textbf{GPU A100}: NVIDIA A100-SXM4 (80 GB)
\item \textbf{Estimated GPU H100~\cite{luo2024benchmarking}}: NVIDIA H100 (80 GB)
\end{enumerate}

We develop an in-house cycle-accurate simulator to model RAPID-Graph’s functionality. PCM arrays and peripheral circuits are modeled with NeuroSim~\cite{peng2020dnn+} for device-level accuracy. We synthesized the RTL in System Verilog using Synopsys Design Compiler with a 40\,nm CMOS PDK at 500\,MHz, including custom permutation unit, min-comparator tree, and controller; all results are scaled to 14 nm using models~\cite{stillmaker2017scaling}.

\subsection{Hardware Configurations}

Each PCM-FW tile integrates a permutation unit for FW, while each PCM-MP tile includes a min-comparator tree for MP. As shown in Table~\ref{tab:pcm-breakdown}, $82$\% of unit area stems from peripheral circuits (sense amplifiers, 1-bit comparators, WL/SL drivers, decoders, D latch buffers), while compute units contribute negligible overhead, supporting dense integration and efficient APSP execution. PIM logic integration lowers PCM density but removes costly data movement, delivering higher performance and efficiency.

System-level supporting components contribute moderate power and area: HBM3 (16\,GB) adds $8.6$\,W and $121\,mm^2$~\cite{park2024attacc}; FeNAND (16\,TB) adds $6.4$\,W across $3000\,mm^2$; the SM2508 controller adds $3.5$\,W within a $225$\,mm$^2$ BGA package. The total power of $\sim18.5$\,W remains significantly lower than high-end GPUs such as the NVIDIA H100, which consumes up to $700$\,W under peak workloads~\cite{luo2024benchmarking} with higher latency.



\begin{table}[t]
\caption{Area and Power Breakdown per PCM Unit}
\label{tab:pcm-breakdown}
\centering
\resizebox{\columnwidth}{!}{%
\begin{tabular}{|l|cc|cc|}
\hline
\textbf{Component} & \multicolumn{2}{c|}{\textbf{PCM-FW}} & \multicolumn{2}{c|}{\textbf{PCM-MP}} \\
\cline{2-5}
& \textbf{Area ($\mu$m$^2$)} & \textbf{Power (mW)} & \textbf{Area ($\mu$m$^2$)} & \textbf{Power (mW)} \\
\hline
PCM Subarray        & 3288 (13.80\%)  & 557 (80.64\%)     & 3288 (13.61\%)  & 557 (80.61\%) \\
Permutation Unit    & 917.3 (3.85\%)  & 0.586 (0.08\%)    & -               & -             \\
Min Comparator      & -               & -                 & 1268 (5.25\%)   & 0.684 (0.10\%) \\
Controller          & 5.94 (0.02\%)   & 0.00126 (<0.01\%) & 5.94 (0.02\%)   & 0.00126 (<0.01\%) \\
Others              & 19610 (82.33\%) & 133.29 (19.28\%)  & 19610 (81.12\%) & 133.29 (19.29\%) \\
\hline
\textbf{Total}      & 23821.24 (100\%) & 690.88 (100\%)   & 24171.94 (100\%) & 690.98 (100\%) \\
\hline
\end{tabular}%
}
\end{table}

\subsection{RAPID-Graph Performance and Efficiency}

\subsubsection{Speedup and Energy Efficiency}

\begin{figure}[t]
    \centering
    \includegraphics[width=0.8\linewidth]{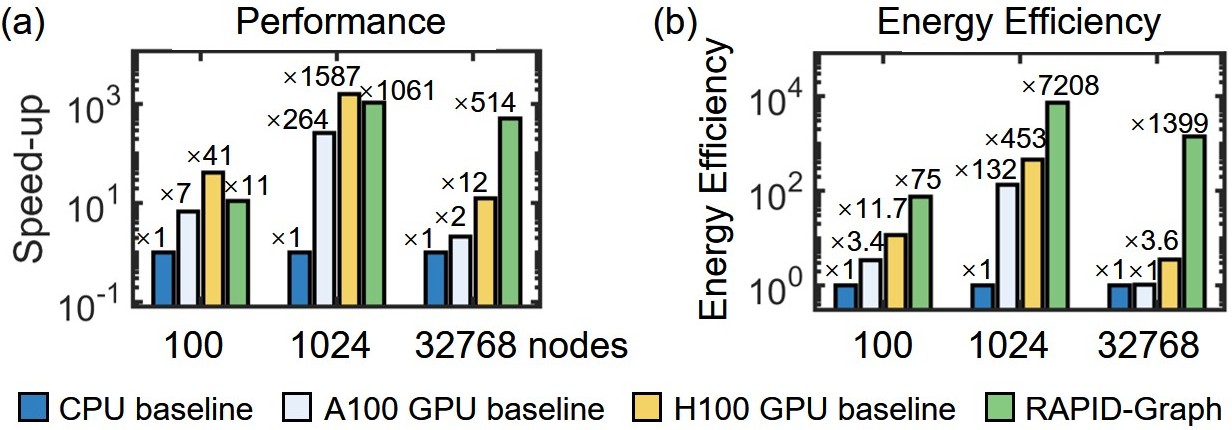}
    \caption{RAPID-Graph vs. CPU, A100 GPU and H100 GPU baselines across graph sizes (a) Speedup (b) Energy efficiency}
    \label{fig:baseline}
\end{figure}

We compare RAPID-Graph against CPU and GPU (A100 and estimated H100) baselines on graphs with 100, 1\,024, and 32\,768 nodes synthesized using NiemaGraphGen~\cite{moshiri2022niemagraphgen}. These sizes avoid memory bottlenecks on single-node hardware. Fig.~\ref{fig:baseline} shows large gains in speed and energy. At 1\,024 nodes, RAPID-Graph delivers $1\,061\times$ speedup and $7\,208\times$ energy efficiency over CPU. At 32\,768 nodes, it outperforms H100 by $42.8\times$ in speed and $392\times$ in energy. RAPID-Graph's performance gains grow with graph size due to enhanced parallelism and in-memory execution. This performance gap widens dramatically with graph size because conventional systems are overwhelmed by $O(n^3)$ data movement. This massive data transfer saturates memory bandwidth that RAPID-Graph inherently avoids.

\begin{figure}[t]
    \centering
    \includegraphics[width=0.8\linewidth]{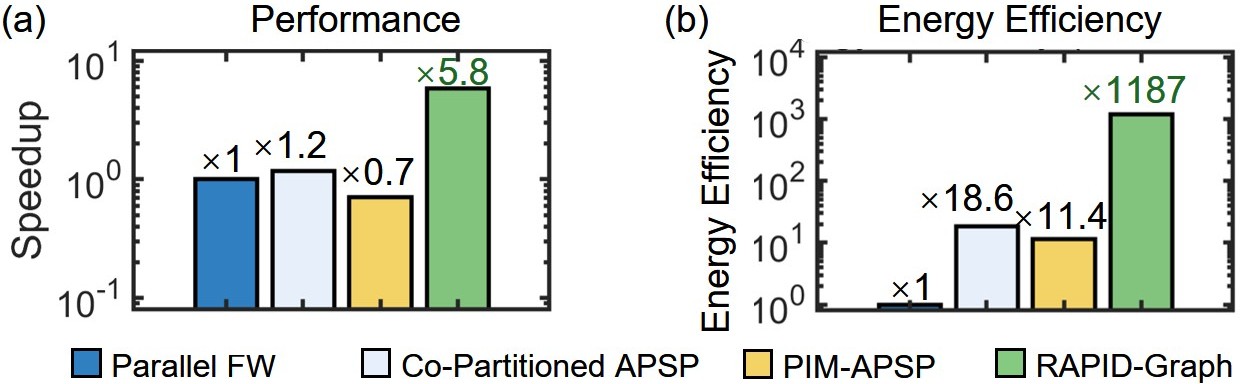}
    \caption{RAPID-Graph vs. PIM-APSP~\cite{madhavan2021temporal}, Partitioned APSP~\cite{djidjev2015all}, and Co-Parallel APSP~\cite{sao2021scalable} running APSP on OGBN-Products dataset~\cite{chiang2019cluster} (a) Speedup (b) Energy efficiency}
    \label{fig:existingwork}
\end{figure}

We also compare RAPID-Graph with SOTA PIM method PIM-APSP~\cite{madhavan2021temporal} and SOTA GPU clusters methods Partitioned APSP~\cite{djidjev2015all} and Co-Parallel APSP~\cite{sao2021scalable}. On OGBN-products (2.5M nodes) dataset~\cite{chiang2019cluster}, we estimate their performance from reported scaling trends. As shown in Fig.~\ref{fig:existingwork}, RAPID-Graph outperforms both, achieving $5.8\times$ speedup over Co-Parallel APSP and $1\,186\times$ energy savings over Partitioned APSP. While PIM-APSP improves energy efficiency by $11.4\times$, it slows down performance to $0.7\times$ of the baseline. The advantage of RAPID-Graph comes from removing inter-GPU communication, the dominant limiter at large-scale, multi-node APSP solutions.


\subsubsection{Scalability}

\begin{figure} [t]
    \centering
    \includegraphics[width=1\linewidth]{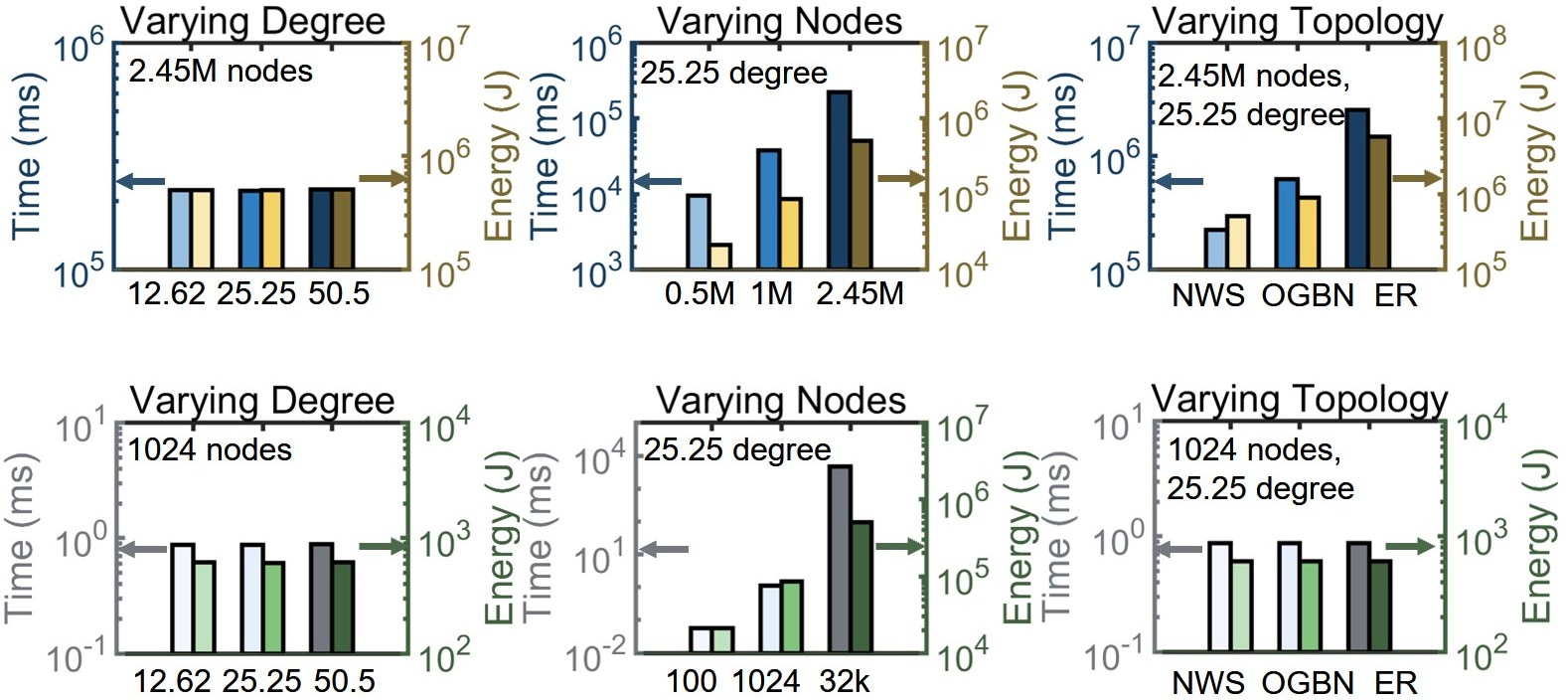}
    \caption{Scalability of RAPID-Graph (top row) and H100 GPU (bottom row) (a,d) Degree sweep at fixed size (b,e) Size sweep at degree 25.25 (c,f) Topology clustered (NWS), real (OGBN), and random (ER) at fixed size and degree}
    \label{fig:scalability}
\end{figure}

Fig.~\ref{fig:scalability} compares the scalability of RAPID-Graph and the H100 GPU baseline across graph degree, size, and topology. In Fig.~\ref{fig:scalability}(a,d), both systems maintain stable performance across degree, suggesting edge count has limited effect on exact APSP when memory is sufficient. In Fig.~\ref{fig:scalability}(b,e), RAPID-Graph scales linearly to 2.45M nodes, while H100 exhibits rising latency and superlinear energy growth beyond $10^3$ nodes due to communication overhead. H100 is limited by the $O(n^2)$ distance matrix overwhelms its caches, whereas RAPID-Graph's in-situ computation preserves locality. In Fig.~\ref{fig:scalability}(c,f), RAPID-Graph achieves better efficiency on clustered and real-world graphs than on random ones, benefiting from structural locality and fewer partitioning boundaries, while H100 remains largely topology-insensitive. This topology-awareness is a direct result of our partitioning algorithm, as clustered graphs like NWS produce smaller boundary sets, reducing the workload of the computationally dominant boundary-graph APSP step. RAPID-Graph provides competitive scalability without the immense hardware cost and overhead of GPU clusters. Partitioned-FW~\cite{djidjev2015all} uses 2,560 GPUs for a 1.9M-node graph but hits synchronization and memory walls. Co-ParallelFW~\cite{sao2021scalable} achieves only 45\% weak-scaling efficiency on a 300K-node graph. Overall, RAPID-Graph avoids the complexity, synchronization, and energy overhead of distributed GPU clusters.


\section{Conclusion}
\label{sec:conclusion}
We present RAPID-Graph, the first software–hardware co-designed PIM system for exact APSP. A recursive partitioner maps subgraphs into PCM arrays, enabling fully parallel FW and MP kernels in-place. The 2.5D stack connects two PCM dies, 16\,GB 8-Hi HBM3, and 16\,TB FeNAND via a $2\,$Tb/s UCIe interposer, sustaining high on-package bandwidth within a $2$\,kW power envelope. On real and synthetic graphs, RAPID-Graph achieves up to $1\,061\times$ speedup and $7\,208\times$ energy efficiency over an Intel i7-11700K CPU, and $42.8\times$ / $392\times$ over an NVIDIA H100 GPU. On the 2.45M-node OGBN-Products graph, it outperforms SOTA GPU clusters by $5.8\times$ in runtime and $1\,186\times$ in energy efficiency, while also showing an $8.3\times$ speedup and $104\times$ energy gain over prior PIM-based solutions. It maintains flat performance across a $4\times$ degree sweep, scales linearly up to 2.45M nodes, and benefits from locality-aware partitioning on clustered and real-world topologies. RAPID-Graph delivers cluster-scale APSP performance at a fraction of the cost, power, and runtime.

\section*{Acknowledgment}
ChatGPT-5 was used to check grammar and enhance clarity of the text.
This work was supported in part by PRISM and CoCoSys—centersand CoCoSys—centers in JUMP 2.0, an SRC program sponsored by DARPA (SRC grant number - 2023-JU-3135). This work was also supported by NSF grants \#2003279, \#1911095, \#2112167, \#2052809, \#2112665, \#2120019, \#2211386.


\balance

\bibliographystyle{IEEEtran}
\bibliography{Ref}

@article{putriani2024comparison,
  title={Comparison of the Dijkstra’s Algorithm and the Floyd-Warshall’s Algorithm to Determine the Shortest Path between Hospitals in Several Cities in Lampung Province},
  author={Putriani, Alenia Daynur and Chasanah, Siti Laelatul and others},
  journal={Journal of Mathematical Sciences and Optimization},
  volume={1},
  number={2},
  pages={44--52},
  year={2024}
}

@article{aktacs2024speeding,
  title={Speeding up the Passenger Assignment Problem in transit planning by the Direct Link Network representation},
  author={Akta{\c{s}}, Dilay and Vermeir, Evert and Vansteenwegen, Pieter},
  journal={Computers \& Operations Research},
  volume={167},
  pages={106647},
  year={2024},
  publisher={Elsevier}
}

@article{ma2024using,
  title={On using Floyd-Warshall under uncertainty for Influence Maximization in Instagram social network: A case study of Indonesian FnB unicorn company},
  author={Ma'ady, Mochamad Nizar Palefi and Syahda, Tabina Shafa Nabila and Rizqi, Annisa Fairuz and Ratna, Maharani Citra Adi},
  journal={Procedia Computer Science},
  volume={234},
  pages={164--171},
  year={2024},
  publisher={Elsevier}
}

@article{madhavan2021temporal,
  title={Temporal state machines: Using temporal memory to stitch time-based graph computations},
  author={Madhavan, Advait and Daniels, Matthew W and Stiles, Mark D},
  journal={JETC},
  volume={17},
  number={3},
  pages={1--27},
  year={2021},
  publisher={ACM New York, NY}
}

@article{wan2021survey,
  title={A survey of fpga-based robotic computing},
  author={Wan, Zishen and Yu, Bo and Li, Thomas Yuang and Tang, Jie and Zhu, Yuhao and Wang, Yu and Raychowdhury, Arijit and Liu, Shaoshan},
  journal={IEEE Circuits and Systems Magazine},
  volume={21},
  number={2},
  pages={48--74},
  year={2021},
  publisher={IEEE}
}

@article{simas2021distance,
  title={The distance backbone of complex networks},
  author={Simas, Tiago and Correia, Rion Brattig and Rocha, Luis M},
  journal={Journal of Complex Networks},
  volume={9},
  number={6},
  pages={cnab021},
  year={2021},
  publisher={Oxford University Press}
}

@inproceedings{yang2023fast,
  title={Fast all-pairs shortest paths algorithm in large sparse graph},
  author={Yang, Shaofeng and Liu, Xiandong and Wang, Yunting and He, Xin and Tan, Guangming},
  booktitle={Proceedings of the 37th International Conference on Supercomputing},
  pages={277--288},
  year={2023}
}

@article{djidjev2015all,
  title={All-Pairs Shortest Path algorithms for planar graph for GPU-accelerated clusters},
  author={Djidjev, Hristo and Chapuis, Guillaume and Andonov, Rumen and Thulasidasan, Sunil and Lavenier, Dominique},
  journal={Journal of Parallel and Distributed Computing},
  volume={85},
  pages={91--103},
  year={2015},
  publisher={Elsevier}
}

@inproceedings{sao2021scalable,
  title={Scalable all-pairs shortest paths for huge graphs on multi-GPU clusters},
  author={Sao, Piyush and Lu, Hao and Kannan, Ramakrishnan and Thakkar, Vijay and Vuduc, Richard and Potok, Thomas},
  booktitle={Proceedings of the 30th International Symposium on High-Performance Parallel and Distributed Computing},
  pages={121--131},
  year={2021}
}

@INPROCEEDINGS{8327035,
  author={Song, Linghao and Zhuo, Youwei and Qian, Xuehai and Li, Hai and Chen, Yiran},
  booktitle={2018 HPCA}, 
  title={GraphR: Accelerating Graph Processing Using ReRAM}, 
  year={2018},
  volume={},
  number={},
  pages={531-543},
  doi={10.1109/HPCA.2018.00052}}

@INPROCEEDINGS{7920847,
  author={Nai, Lifeng and Hadidi, Ramyad and Sim, Jaewoong and Kim, Hyojong and Kumar, Pranith and Kim, Hyesoon},
  booktitle={2017 IEEE International Symposium on High Performance Computer Architecture (HPCA)}, 
  title={GraphPIM: Enabling Instruction-Level PIM Offloading in Graph Computing Frameworks}, 
  year={2017},
  volume={},
  number={},
  pages={457-468},
  doi={10.1109/HPCA.2017.54}}

@inproceedings{sun2017graphh,
  title={Graphh: High performance big graph analytics in small clusters},
  author={Sun, Peng and Wen, Yonggang and Duong, Ta Nguyen Binh and Xiao, Xiaokui},
  booktitle={2017 CLUSTER},
  pages={256--266},
  year={2017},
  organization={IEEE}
}

@inproceedings{lin202536,
  title={36.1 A 32Gb/s 10.5 Tb/s/mm 0.6 pJ/b UCIe-Compliant Low-Latency Interface in 3nm Featuring Matched-Delay for Dynamic Clock Gating},
  author={Lin, Mu-Shan and Tsai, Chien-Chun and Li, Shenggao and Chen, Wei-Chih and Huang, Wen-Hung and Chen, Yu-Chi and Huang, Yu-Jie and Drake, Alan and Wen, Chin-Hua and Ranucci, Paul and others},
  booktitle={2025 IEEE International Solid-State Circuits Conference (ISSCC)},
  volume={68},
  pages={586--588},
  year={2025},
  organization={IEEE}
}

@article{wu2024novel,
  title={Novel nanocomposite-superlattices for low energy and high stability nanoscale phase-change memory},
  author={Wu, Xiangjin and Khan, Asir Intisar and Lee, Hengyuan and Hsu, Chen-Feng and Zhang, Huairuo and Yu, Heshan and Roy, Neel and Davydov, Albert V and Takeuchi, Ichiro and Bao, Xinyu and others},
  journal={Nature Communications},
  volume={15},
  number={1},
  pages={13},
  year={2024},
  publisher={Nature Publishing Group UK London}
}

@article{karypis1998multilevelk,
  title={Multilevelk-way partitioning scheme for irregular graphs},
  author={Karypis, George and Kumar, Vipin},
  journal={Journal of Parallel and Distributed computing},
  volume={48},
  number={1},
  pages={96--129},
  year={1998},
  publisher={Elsevier}
}

@inproceedings{gupta2018felix,
  title={Felix: Fast and energy-efficient logic in memory},
  author={Gupta, Saransh and Imani, Mohsen and Rosing, Tajana},
  booktitle={2018 IEEE/ACM International Conference on Computer-Aided Design (ICCAD)},
  pages={1--7},
  year={2018},
  organization={IEEE}
}

@article{wong2010phase,
  title={Phase change memory},
  author={Wong, H-S Philip and Raoux, Simone and Kim, SangBum and Liang, Jiale and Reifenberg, John P and Rajendran, Bipin and Asheghi, Mehdi and Goodson, Kenneth E},
  journal={Proceedings of the IEEE},
  volume={98},
  number={12},
  pages={2201--2227},
  year={2010},
  publisher={IEEE}
}

@article{floyd1962algorithm,
  title={Algorithm 97: shortest path},
  author={Floyd, Robert W},
  journal={Communications of the ACM},
  volume={5},
  number={6},
  pages={345--345},
  year={1962},
  publisher={ACM New York, NY, USA}
}

@incollection{dijkstra2022note,
  title={A note on two problems in connexion with graphs},
  author={Dijkstra, Edsger W},
  booktitle={Edsger Wybe Dijkstra: his life, work, and legacy},
  pages={287--290},
  year={2022}
}

@inproceedings{chiang2019cluster,
  title={Cluster-gcn: An efficient algorithm for training deep and large graph convolutional networks},
  author={Chiang, Wei-Lin and Liu, Xuanqing and Si, Si and Li, Yang and Bengio, Samy and Hsieh, Cho-Jui},
  booktitle={Proceedings of the 25th ACM SIGKDD international conference on knowledge discovery \& data mining},
  pages={257--266},
  year={2019}
}

@article{peng2020dnn+,
  title={DNN+ NeuroSim V2. 0: An end-to-end benchmarking framework for compute-in-memory accelerators for on-chip training},
  author={Peng, Xiaochen and Huang, Shanshi and Jiang, Hongwu and Lu, Anni and Yu, Shimeng},
  journal={IEEE Transactions on Computer-Aided Design of Integrated Circuits and Systems},
  volume={40},
  number={11},
  pages={2306--2319},
  year={2020},
  publisher={IEEE}
}

@article{moshiri2022niemagraphgen,
  title={NiemaGraphGen: A memory-efficient global-scale contact network simulation toolkit},
  author={Moshiri, Niema},
  journal={GIGAbyte},
  volume={2022},
  pages={gigabyte37},
  year={2022}
}

@article{watts1998collective,
  title={Collective dynamics of ‘small-world’networks},
  author={Watts, Duncan J and Strogatz, Steven H},
  journal={nature},
  volume={393},
  number={6684},
  pages={440--442},
  year={1998},
  publisher={Nature Publishing Group}
}

@article{erdds1959random,
  title={On random graphs I},
  author={ERDdS, P and R\&wi, A},
  journal={Publ. math. debrecen},
  volume={6},
  number={290-297},
  pages={18},
  year={1959}
}

@inproceedings{luo2024benchmarking,
  title={Benchmarking and dissecting the nvidia hopper gpu architecture},
  author={Luo, Weile and Fan, Ruibo and Li, Zeyu and Du, Dayou and Wang, Qiang and Chu, Xiaowen},
  booktitle={2024 IPDPS},
  pages={656--667},
  year={2024},
  organization={IEEE}
}

@inproceedings{ahn2015scalable,
  title={A scalable processing-in-memory accelerator for parallel graph processing},
  author={Ahn, Junwhan and Hong, Sungpack and Yoo, Sungjoo and Mutlu, Onur and Choi, Kiyoung},
  booktitle={Proceedings of the 42nd annual international symposium on computer architecture},
  pages={105--117},
  year={2015}
}

@inproceedings{zhang2018graphp,
  title={GraphP: Reducing communication for PIM-based graph processing with efficient data partition},
  author={Zhang, Mingxing and Zhuo, Youwei and Wang, Chao and Gao, Mingyu and Wu, Yongwei and Chen, Kang and Kozyrakis, Christos and Qian, Xuehai},
  booktitle={2018 IEEE International Symposium on High Performance Computer Architecture (HPCA)},
  pages={544--557},
  year={2018},
  organization={IEEE}
}

@inproceedings{zhuo2019graphq,
  title={Graphq: Scalable pim-based graph processing},
  author={Zhuo, Youwei and Wang, Chao and Zhang, Mingxing and Wang, Rui and Niu, Dimin and Wang, Yanzhi and Qian, Xuehai},
  booktitle={Proceedings of the 52nd Annual IEEE/ACM International Symposium on Microarchitecture},
  pages={712--725},
  year={2019}
}

@INPROCEEDINGS{7551379,
  author={Shafiee, Ali and Nag, Anirban and Muralimanohar, Naveen and Balasubramonian, Rajeev and Strachan, John Paul and Hu, Miao and Williams, R. Stanley and Srikumar, Vivek},
  booktitle={2016 ACM/IEEE 43rd Annual International Symposium on Computer Architecture (ISCA)}, 
  title={ISAAC: A Convolutional Neural Network Accelerator with In-Situ Analog Arithmetic in Crossbars}, 
  year={2016},
  volume={},
  number={},
  pages={14-26},
  doi={10.1109/ISCA.2016.12}}

@article{shi2018graph,
  title={Graph processing on GPUs: A survey},
  author={Shi, Xuanhua and Zheng, Zhigao and Zhou, Yongluan and Jin, Hai and He, Ligang and Liu, Bo and Hua, Qiang-Sheng},
  journal={CSUR},
  volume={50},
  number={6},
  pages={1--35},
  year={2018},
  publisher={ACM New York, NY, USA}
}

@article{xie2025gpu,
  title={GPU Architectures in Graph Analytics: A Comparative Experimental Study},
  author={Xie, Peichen and Zheng, Zhigao and Zhou, Yongluan and Xiu, Yang and Liu, Hao and Yang, Zhixiang and Zhang, Yu and Du, Bo},
  year={2025}
}

@INPROCEEDINGS{10873526,
  author={Lee, Jae-Gil and Koo, Won-Tae and Lee, Geonhui and Kim, Jihun and Lee, Woocheol and Lee, Hyung Dong and Yoon, Sunghyun and Hong, Sung-In and Kang, In-Ku and Kim, Joongsik and Choi, Hyejung and Kim, Soo Gil and Lee, Seho and Yi, Jaeyun and Cha, Seon Yong},
  booktitle={2024 IEEE International Electron Devices Meeting (IEDM)}, 
  title={Analog Computation in Ultra-High Density 3D FeNAND for TB-level Hyperscale AI Models}, 
  year={2024},
  volume={},
  number={},
  pages={1-4},
  doi={10.1109/IEDM50854.2024.10873526}}

@article{stillmaker2017scaling,
  title={Scaling equations for the accurate prediction of CMOS device performance from 180 nm to 7 nm},
  author={Stillmaker, Aaron and Baas, Bevan},
  journal={Integration},
  volume={58},
  pages={74--81},
  year={2017},
  publisher={Elsevier}
}

@article{he2023ti,
  title={Ti/HfO2-based RRAM with superior thermal stability based on self-limited TiOx},
  author={He, Huikai and Tan, Yixin and Lee, Choonghyun and Zhao, Yi},
  journal={Electronics},
  volume={12},
  number={11},
  pages={2426},
  year={2023},
  publisher={MDPI}
}

@inproceedings{nguyen2025low,
  title={Low write power and Field-free sub-ns write speed SOT-MRAM cell with Design Technology of Canted SOT structure and Magnetic Anisotropy for NVM},
  author={Nguyen, TVA and Naganuma, H and Honjo, H and Sato, Y and Tanigawa, T and Ikeda, S and Endoh, T},
  booktitle={2025 IEEE International Memory Workshop (IMW)},
  pages={1--4},
  year={2025},
  organization={IEEE}
}

@article{park2024phase,
  title={Phase-change memory via a phase-changeable self-confined nano-filament},
  author={Park, See-On and Hong, Seokman and Sung, Su-Jin and Kim, Dawon and Seo, Seokho and Jeong, Hakcheon and Park, Taehoon and Cho, Won Joon and Kim, Jeehwan and Choi, Shinhyun},
  journal={Nature},
  volume={628},
  number={8007},
  pages={293--298},
  year={2024},
  publisher={Nature Publishing Group UK London}
}

@article{yun2024optimization,
  title={The Optimization of Program Operation for Low Power Consumption in 3D Ferroelectric (Fe)-NAND Flash Memory},
  author={Yun, Myeongsang and Lee, Gyuhyeon and Ryu, Gyunseok and Kim, Hyoungsoo and Kang, Myounggon},
  journal={Electronics},
  volume={13},
  number={2},
  pages={316},
  year={2024},
  publisher={MDPI}
}

@inproceedings{park2024attacc,
  title={AttAcc! Unleashing the power of PIM for batched transformer-based generative model inference},
  author={Park, Jaehyun and Choi, Jaewan and Kyung, Kwanhee and Kim, Michael Jaemin and Kwon, Yongsuk and Kim, Nam Sung and Ahn, Jung Ho},
  booktitle={Proceedings of the 29th ACM International Conference on Architectural Support for Programming Languages and Operating Systems, Volume 2},
  pages={103--119},
  year={2024}
}

\end{document}